\documentclass[letterpaper,twocolumn,10pt]{article}
\usepackage{usenix-2020-09}

\usepackage{hyperref}
\usepackage{amsmath}
\usepackage{cleveref}
\usepackage{adjustbox}
\usepackage{mathtools}
\usepackage{dsfont}
\usepackage{xcolor}
\usepackage{amsthm}

\usepackage{tikz}
\usepackage{colortbl}

\usepackage{macros}

\DeclareSpecialMathOperator{\bits}{bits}
\DeclareSpecialMathOperator{\size}{size}
\DeclareSpecialMathOperator{\seq}{seq}
\DeclareSpecialMathOperator{\timecert}{timecert}
\DeclareSpecialMathOperator{\commit}{digest}
\DeclareSpecialMathOperator{\certify}{certify}
\DeclareSpecialMathOperator{\sparsecommit}{iterative\_digest}
\DeclareSpecialMathOperator{\sparsecertify}{iterative\_certify}
\DeclareSpecialMathOperator{\verify}{verify}
\DeclareSpecialMathOperator{\gcommit}{digest\_vertex}
\DeclareSpecialMathOperator{\dock}{commit}
\DeclareSpecialMathOperator{\gcertify}{certificate\_vertices}
\DeclareSpecialMathOperator{\certificatepool}{certificate\_pool}
\DeclareSpecialMathOperator{\commitpool}{digest\_pool}
\DeclareSpecialMathOperator{\nextroot}{next\_root}
\DeclareSpecialMathOperator{\nextpower}{next\_power}
\DeclareSpecialMathOperator{\nextp}{next_{2}}
\DeclareSpecialMathOperator{\prevp}{prev_{2}}
\DeclareSpecialMathOperator{\nexto}{next_{3}}
\DeclareSpecialMathOperator{\prevo}{prev_{3}}
\DeclareSpecialMathOperator{\nextv}{next}
\DeclareSpecialMathOperator{\maxpowop}{maxpow}
\newcommand{\maxpow}[2]{\maxpowop_{#1}(#2)}

\begin{document}
%-------------------------------------------------------------------------------

\usetikzlibrary{calc}
\usetikzlibrary{fadings}
\usetikzlibrary{decorations.pathmorphing}

\definecolor{cCert}{RGB}{255, 201, 143}
\definecolor{cCertText}{RGB}{212, 110, 2}
\definecolor{cCertPathText}{RGB}{127, 3, 252}
\definecolor{cCertPool}{RGB}{201, 222, 255}
\definecolor{cCertPoolText}{RGB}{65, 116, 196}
\definecolor{cSkip}{RGB}{0, 120, 68}
\definecolor{myOrange}{rgb}{1.0, 0.66, 0.07}
\colorlet{myRed}{red!90!black}
\definecolor{myBlue}{rgb}{0.25, 0.0, 1.0}
\colorlet{cDeemph}{black!50!white}

\definecolor{cProofPath1}{rgb}{0.25, 0.0, 1.0}
\definecolor{cProofPath2}{rgb}{0.4, 0.0, 0.80}

% \definecolor{cCertPath}{RGB}{223, 191, 255}
\colorlet{cCertPath}{cProofPath1!25!white}
\colorlet{cCertPathText}{cProofPath1!90!black}
\colorlet{cProofPathText1}{cProofPath1!90!black}
\colorlet{cProofPathText2}{cProofPath2!90!black}

\definecolor{cPoolPath1}{rgb}{0.0, 0.3, 1.0}
\definecolor{cPoolPath2}{rgb}{0.0, 0.6, 0.9}
\definecolor{cPoolPath3}{rgb}{0.0, 0.8, 0.7}
\definecolor{cPoolPath4}{rgb}{0.4, 0.8, 0.0}
\definecolor{cPoolPath5}{RGB}{252, 104, 166}

\colorlet{cPoolPath}{cPoolPath1!25!white}
\definecolor{cPoolPathText1}{rgb}{0.0, 0.3, 0.8}
\colorlet{cPoolPathText2}{cPoolPath2!80!black}
\colorlet{cPoolPathText3}{cPoolPath3!80!black}
\colorlet{cPoolPathText4}{cPoolPath4!80!black}
\colorlet{cPoolPathText5}{cPoolPath5!80!black}

\colorlet{cStart}{cProofPath1!90!black}
\definecolor{cMiddle}{RGB}{212, 110, 2}

\definecolor{cPrevCore}{RGB}{170, 230, 198}
\definecolor{cPrevCoreText}{RGB}{10, 99, 52}

\colorlet{myRedText}{myRed!80!black}

\definecolor{cCoreFourth}{RGB}{240, 143, 255}

\colorlet{cLightGrey}{black!10!white}

\begin{tikzfadingfrompicture}[name=faderight]   
    \clip (0,0) rectangle (2,2);   
    \shade[left color=black,right color=white] (0,0) rectangle (1.7,2);
    \shade[left color=white,right color=white] (1.7,0) rectangle (2.1,2);                
\end{tikzfadingfrompicture}

\tikzset{
    myedge/.style={
        arrows=->
    },
    edgePredecessor/.style={
        arrows=->,
        color=cDeemph
    },
    edgeSkip/.style={
        arrows=->,
        color=cSkip,
        very thick
    },
    skipEdge/.style={
        arrows=->,
    },
    emphedge/.style={
        arrows=->,
        color=myRed,
        ultra thick
    },
    edgeParent/.style={
        arrows=->,
        color=cDeemph,
    },
    nodeItem/.style={
        color=cDeemph,
    },
    certPathV/.style={
        fill=cCertPath,
        rounded corners=5
    },
    certPathE/.style={
        preaction={draw,line width=6,cCertPath,line cap=round,arrows=-}
    },
    certV/.style={
        fill=cCert,
    },
    certPoolV/.style={
        fill=cPoolPath,
        rounded corners=5
    },
    certPoolE/.style={
        preaction={draw,line width=6,cPoolPath,line cap=round,arrows=-}
    },
    toProve/.style={
        font=\boldmath,
    },
    digest/.style={
        font=\boldmath,
    },
    emphnode/.style={
        fill=myRed,
        shape=circle,
        minimum size=0.02cm,
    },
    deemphnode/.style={
        color=black!10!white
    },
    deemphedge/.style={
        color=black!10!white
    },
    prevcorestart/.style={
        fill=cPrevCore,
        shape=circle,
        minimum size=0.05cm,
    },
    corestart/.style={
        fill=cCertPath,
        shape=circle,
        minimum size=0.05cm,
    },
    coremiddle/.style={
        fill=cCert,
        shape=circle,
        minimum size=0.05cm,
    },
    corefourth/.style={
        fill=cCoreFourth,
        shape=circle,
        minimum size=0.05cm,
    },
    coreother/.style={
        fill=black!6!white,
        shape=circle,
        minimum size=0.02cm,
    },
    codevariablepi/.style={
    },
    codevariablepipp/.style={
    },
    codevariablevi/.style={
    },
    separatingline/.style={
        gray,
        dashed
    },
    stylerepairededge/.style={
        preaction={draw,line width=10,cCert,line cap=round}
    },
    styleinvalidedge/.style={
        preaction={draw,line width=10,cCert,line cap=round}
    },
    stylepinode/.style={
        fill=colorrepairededge
    },
    highlightedge/.style={
        preaction={draw,line width=10,cCertPath,line cap=round}
    },
    highlightnode/.style={
        fill=cCertPath,
        rounded corners=8
    },
    squigglyline/.style={decorate, decoration={snake, segment length=6mm, amplitude=0.5mm}},
}

%don't want date printed
\date{}

\title{\Large \bf Better Prefix Authentication}

\author{
{\rm Aljoscha Meyer}\\
Technical University Berlin
}
% \author{
% {\rm Anonymous Authors}\\
% Anonymous Affiliations
% }

\maketitle

%-------------------------------------------------------------------------------
\begin{abstract}
%-------------------------------------------------------------------------------
We present new schemes for solving prefix authentication and secure relative timestamping. By casting a new light on antimonotone linking schemes, we improve upon the state of the art in prefix authentication, and in timestamping with rounds of bounded length. Our designs can serve as more efficient alternatives to certificate transparency logs.
\end{abstract}

%-------------------------------------------------------------------------------
\section{Introduction}
%-------------------------------------------------------------------------------

% 13 typeset pages, excluding bibliography and well-marked appendices

Prefix authentication~\cite{meyer2023prefix} asks to annotate finite sequences with metadata so that one can cryptographically prove when some sequence is a prefix of another. It generalizes \textit{tamper-evident logging}~\cite{crosby2009efficient} and \textit{secure timestamping}~\cite{haber1990time}.

In recent years, \textit{certificate transparency}~\cite{rfc9162} and related transparency logging schemes~\cite{fahl2014hey}\cite{nikitin2017chainiac}\cite{al2018contour} have garnered significant attention. There has been little progress on improving the efficiency of the underlying prefix authentication schemes however.

We provide the first prefix authentication schemes that asymptotically (and practically) outperform the certificate transparency logs, both in the amount of required metadata and in the size of the data that certifies that some string is a prefix of another.

\Cref{related_work} summarizes related work, \cref{preliminaries} gives the preliminaries to make sense of what follows. \Cref{lssl2} introduces our core idea of using slightly modified skip lists, exemplified by a \textit{binary} skip list. \Cref{lssl3} presents the more efficient \textit{ternary} skip list. We discuss the results in \cref{discussion} before concluding in \cref{conclusion}.

Finally, a note on presentation style. We feel like our results could and should have been discovered two decades ago, had the literature on secure timestamping via linking schemes been more accessible to distributed systems practitioners. So we deliberately focus on concrete schemes rather than abstract theory, keep the prose compact and the figures plentiful, and favor intuitive, high-level reasoning over formalisms. Certificate transparency logs are less elegant than linking schemes (and less efficient as we are about to show), so we hope the ideas stick around this time. Those readers who wish for a more rigorous, formal treatment of prefix authentication via hash-labeled graphs can find it in~\cite{meyer2023prefix}.

\section{Related Work}\label{related_work}

Prefix authentication~\cite{meyer2023prefix} unifies several previously disjoint strands of research, such as secure logging~\cite{schneier1999secure}\cite{crosby2009efficient}\cite{pulls2013distributed}, accountable shared storage~\cite{li2004secure}\cite{yumerefendi2007strong}, certificate transparency~\cite{laurie2014certificate}\cite{laurie2014certificateb}\cite{rfc9162}, or data replication~\cite{ogden2017dat}\cite{tarr2019secure}. Our designs fall in the class of \textit{transitive prefix authentication schemes}, more specifically they are \textit{linking schemes}.

Our two linking schemes are almost isomorphic to the \textit{simple antimonotone binary linking scheme}~\cite{buldas1998time} and the \textit{optimal antimonotone binary linking scheme}~\cite{buldas1998new} respectively. The latter paper introduces the \textit{antimonotone graph product} $\otimes$ as an operation that generates all antimonotone binary graphs. While we do not explicitly define or rely on this operator, we occasionally mention it to point the interested reader to the parallels to these works.

The simple and optimal antimonotone binary linking schemes rely on the same underlying graphs as ours, but the way these graphs are then used to  provide prefix authentication is suboptimal (as is their direct transformation into an \textit{anti-monotone $\phi$-scheme}~\cite{buldas2000optimally}).

\textit{Threaded authentication trees}~\cite{buldas2000optimally}, \textit{hypercore}~\cite{ogden2017dat}, and \textit{certificate transparency logs}~\cite{rfc9162} are the most efficient prefix authentication schemes to date. We outperform each of them.

We derive our schemes not from antimonotone theory but from slightly modified, deterministic skip lists~\cite{pugh1990skip}. Chainiac~\cite{nikitin2017chainiac} is a prior approach to prefix authentication based on skip lists, but a significantly less efficient one. Blibech and Gabillon~\cite{blibech2006new} have proposed a skip list timestamping scheme of similar efficiency as ours, but their scheme relies on knowing the maximum length of a timestamping round. As such, it is only applicable to prefix authentication for sequences of bounded length, but not for sequences of arbitrary length.

\section{Preliminaries}\label{preliminaries}

Rather than laying out the general theory of transitive prefix authentication schemes~\cite{meyer2023prefix}, we present a simple subset that captures the schemes we will present. The results from the underlying theories still apply, but we get to significantly simplify our presentation.

We assume basic understanding of cryptographic hash functions~\cite{menezes2018handbook}, and a basic background in graph theory~\cite{west2001introduction}; we consider \textit{directed} graphs exclusively.

When an object contains a secure hash of another object, the second object must have already existed at the creation time of the first object. This property transitively carries over, and forms the basis of our techniques.

We can represent objects that contain hashes of other objects as labeled DAGs, generalizing the well-known Merkle trees~\cite{merkle1989certified}: sinks are labeled with the hashes of some objects of interest (say, the elements of a sequence to authenticate), the other vertices are labeled with the hash of the concatenation of their out-neighbors' labels.

\begin{figure}
  \centering
  \includegraphics{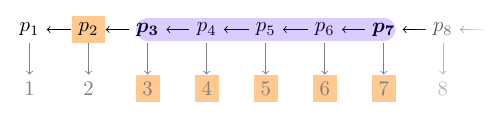}
  \caption{The linear linking scheme. The \textcolor{cCertText}{out-neighborhood} of the \textcolor{cCertPathText}{path from $\gcommit(7)$ to $\gcommit(3)$} determines the label of $\gcommit(7)$: the labels of $p_2$ and $3$ determine the label of $p_3$, this together with the label of $4$ determines the label of $p_4$, and so on.}
  \label{fig_linear}
\end{figure}

\begin{figure*}
  \centering
  \includegraphics{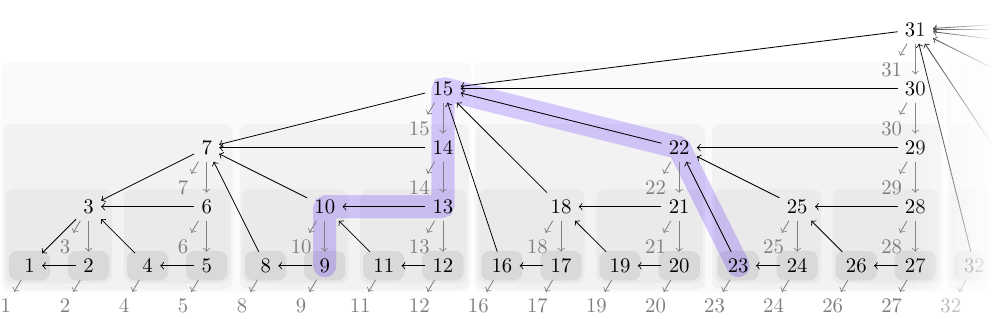}
  \caption{The simple antimonotone binary linking scheme $G_{ls2}$, highlighting the \textcolor{cCertPathText}{shortest path from $23$ to $9$}.}
  \label{fig_abls2}
\end{figure*}

For our purposes, a \defined{linking scheme} is an acyclic graph whose sinks are the natural numbers without zero, in which every sink $n$ has exactly one parent vertex, denoted $\gcommit(n)$, and in which there is a path from every $\gcommit(n + 1)$ to $\gcommit(n)$. \Cref{fig_linear} depicts the simplemost such graph, \cref{fig_abls2} shows the more sophisticated \textit{simple antimonotone binary linking scheme}~\cite{buldas1998time}.

Hash-labeled linking schemes can provide prefix authentication. Let $t$ be a sequence of length $len_t$, then label each sink $n \leq len_t$ with the hash of the $n$-th sequence item. For any $len_s \leq len_t$, $\gcommit(len_s)$ authenticates the prefix $s$ of length $len_s$, since all $n \leq len_s$ are reachable from it and thus influence its hash.

It hence suffices to certify that $\gcommit(len_s)$ is reachable from $\gcommit(len_t)$ in order to prove that $s$ is a prefix of $t$. This \defined{prefix certificate of $s$ and $t$} consist of the labels of the closed out-neighborhood of a (shortest) path from $\gcommit(len_t)$ to $\gcommit(len_s)$: these labels suffice to reconstruct the label of $\gcommit(len_t)$, thus proving that there is indeed a path from $\gcommit(len_t)$ to $\gcommit(len_s)$ (compare \cref{fig_linear}).

In order to evaluate the efficiency of a linking scheme, we ask for a mapping $\fun{\certificatepool}{\Np}{\powerset{V}}$ that maps every number to a set of vertices such that for all $len_s < len_t$ the shortest path from $\gcommit(len_t)$ to $\gcommit(len_s)$ is contained in the intersection $\certificatepool(len_s) \cup \certificatepool(len_t)$. The out-neighborhood of $\certificatepool(n)$ is called the \defined{positional certificate} of $n$.

A particularly interesting class of linking schemes, due to their similarity to our schemes, are the \defined{antimonotone binary linking schemes}~\cite{buldas1998time}\cite{buldas1998new}. In these schemes, the parent of sink $n$ --- denoted $p_n$ --- has an edge to $p_{n - 1}$, and another edge to $p_{\f(n)}$, where $\f$ is some function that satisfies the \defined{antimonotonicity property} $n < m \implies \f(n) \geq \f(m)$. \Cref{fig_abls2} depicts the \defined{simple antimonotone binary linking scheme}~\cite{buldas1998time}, whose graph $G_{ls2}$ is given by the following function $f_2$:

\begin{align*}
  f_2(n) &\defeq \begin{cases}
      n - (2^{k - 1} + 1) &\mbox{if } n = 2^k - 1, k \in \N \\
      n - 2^{g(n)} & \mbox{otherwise}\\
  \end{cases}\\
  g(n) &\defeq \begin{cases}
      k &\mbox{if } n = 2^{k} - 1, k \in \N \\
      g(n - (2^{k - 1} - 1)) & \mbox{if } 2^{k - 1} - 1 < n < 2^{k} - 1, k \in \N.
  \end{cases}\\
\end{align*}

\section{Skip List Linking Schemes Done Right}\label{lssl2}

Let $n, k \in \Nz$, then $\maxpow{k}{n}$ denotes the largest natural number $p$ such that $k^p$ divides $n$.

\begin{figure*}
  \centering
  \includegraphics{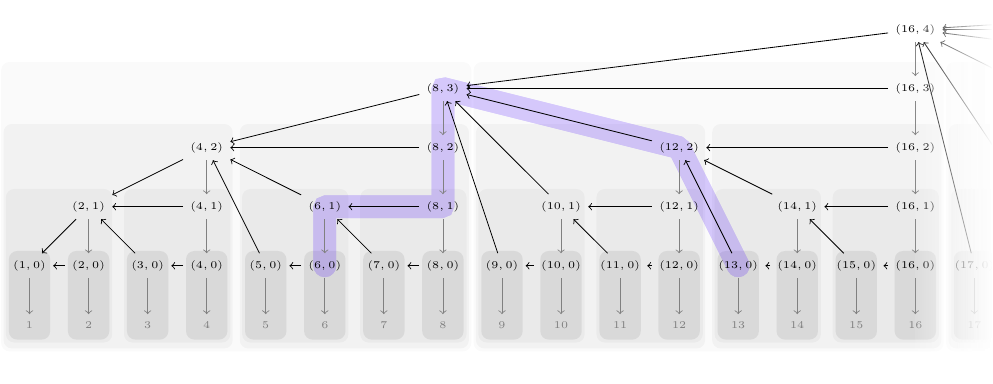}
  \caption{The $SLLS_2$, highlighting the \textcolor{cCertPathText}{shortest path from $(13, 0)$ to $(6, 0)$}.}
  \label{fig_pnls_basic}
\end{figure*}

We now present the \defined{$SLLS_2$} (\defined{binary skip list linking scheme}). Its underlying graph is a skip list with a twist, links that would normally stay in the same layer of the skip list point to the topmost available layer instead (\cref{fig_pnls_basic}):

\begin{align*}
  V_{slls2} &\defeq \set{(n, k) \st \text{$n \in \N, k \in \Nz$ and $2^k \divides n$}}\cup \N,\\
  E_{slls2} &\defeq \Bigl\{\bigl((n, k + 1), (n, k)\bigr)\Bigr\} \cup \Bigl\{\bigl((n, 0), n\bigr)\Bigr\}\\
  &\cup \Bigl\{\bigl((n + 2^k, k), (n, \maxpow{2}{n})\bigr)\Bigr\},\\
  G_{slls2} &\defeq (V_{slls2}, E_{slls2}).
\end{align*}

% Define $\nextv(n)$ as $(n + 1, 0)$ if $\maxpow{2}{n} = 0$ and $(n, 1)$ otherwise. Then define $\gcommit(n) \defeq (n, 0)$, and define $\gcertify(len_s, len_t)$ as the shortest path from $\gcommit(len_t)$ to $\nextv(len_s)$ to obtain a linking scheme.

% For $n$ with $\maxpow{2}{n} = 0$, $(n + 1, 0)$ has an edge to $(n, 0)$. Otherwise, $(n + 1, 0)$ has an edge to some $(n, k)$, which has an edge to $(n, k - 1)$, and so on, down to $(n, 0)$ via $(n, 1)$. In both cases, there is a path from $(n + 1, 0)$ to $(n, 0)$ via $\nextv(n)$, so the definition does indeed yield a linking scheme.

Perhaps surprisingly, the $SLLS_2$ is almost isomorphic to the simple antimonotone binary scheme~\cite{buldas1998time}, compare \cref{fig_pnls_basic} and \cref{fig_abls2}. More precisely, $G_{slls2} \setminus \N$ is isomorphic to $G_{ls2} \setminus \N$ (and both are isomorphic to the limit of the recursive antimonotone product $G_{simple}^{i + 1} \defeq G_{simple}^i \otimes G_{simple}^i \otimes G_1$)~\cite{buldas1998new}.

As a single look at \cref{fig_pnls_basic} and \cref{fig_abls2} leaves little doubt about the isomorphism, we give no further proof beyond stating the bijective function between the vertices. A vertex $(n, k)$ of the $SLLS_2$ maps to the vertex $m$ of $G_{ls2}$, where $m$ is the number of vertices on the longest path from $(n, k)$ to $(1, 0)$. Conversely, a vertex $m$ of $G_{ls2}$ maps to the vertex $(n, k)$ of the $SLLS_2$, where $n$ is the number of edges in the longest path from $m$ to $1$ of the form $\bigl(x, f_2(x)\bigr)$, and $k$ is the number of consecutive edges at the start of the longest path from $m$ to $1$ of the form $(x, x - 1)$. In less formal terms, $(n, k)$ are the x and y coordinates of $m$ when drawing $G_{ls2}$ in the style of \cref{fig_abls2}.

\begin{figure*}
  \centering
  \includegraphics{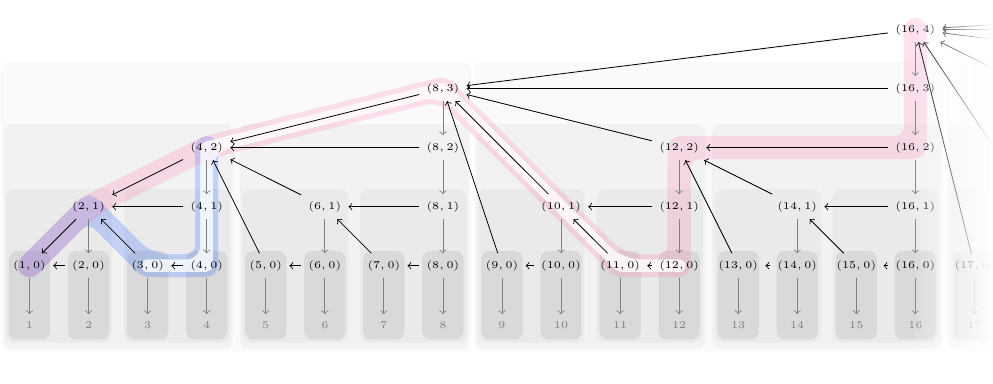}
  \caption{The certificate pools of $3$ and $12$, two numbers from different generations. The vertebra of the generation of the lesser number lies on the spine of the generation of the greater number.}
  \label{fig_poola}
\end{figure*}

\begin{figure*}
  \centering
  \includegraphics{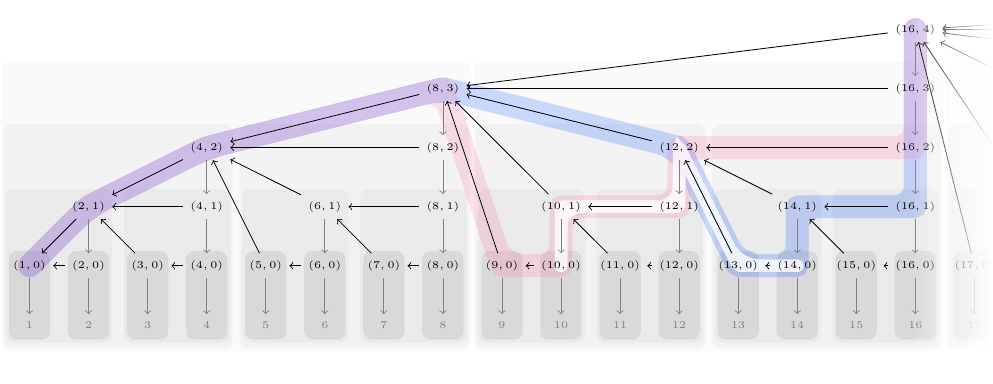}
  \caption{The certificate pools of $10$ and $14$, two numbers from the same generation. Their paths necessarily cross within the generation.}
  \label{fig_poolb}
\end{figure*}

The certificate pools for $SLLS_2$ correspond directly to those of $G_{ls2}$ under the isomorphism. For any number $n$, we say it belongs to \defined{generation} $\ceil{\log_2(n)}$. We say $\bigl(t, \log_2(t)\bigr)$ is the \defined{vertebra} of generation $t$, and the set of vertebrae of all generations up to and including $t$ is the \defined{spine} of generation $t$. The certificate pool of $n$ is the union of the shortest paths from the vertebra of $t$ to $(n, 0)$, from $(n, 0)$ to the vertebra of $t - 1$, and from that vertebra to $(1, 0)$. In the isomorphic antimonotone setting, Buldas et al.~\cite{buldas1998time} have proven that the intersection of two such certificate pool contains the shortest path between the numbers in question, so the construction does indeed yield a valid certificate pool. \Cref{fig_poola} and \cref{fig_poolb} convey the underlying intuition.

The size of the out-neighborhood of the certificate pool of some $n$ is different to that in $G_{ls2}$ however; we claim it is $2 \cdot \ceil{\log_2(n)}$ for all $n > 4$. Let $n \in \N, n > 4$, and let $t$ be the generation of $n$. The shortest path from the vertebra of $t - 1$ to $(1, 0)$ consists of the vertices $(2^{t - 1}, t - 1), (2^{t - 2}, t - 2), \cdots, (2^0, 0)$, for a total of $t$ vertices. Every such vertex has exactly one out-neighbor outside the path: $1$ for $(1, 0)$, and $(x, k - 1)$ for any other $(x, k)$. This brings the out-neighborhood of the certificate pool to a size of at least $\ceil{log_2(n)}$. It remains to show that the out-neighborhood of the shortest path from the vertebra of $t$ to the vertebra of $t - 1$ via $(n, 0)$ has the same size.

We have two different kinds of outgoing edges to consider: those from some $(x, k)$ to $(x, k - 1)$, and those whose first component decreases by some amount. From the antimonotone perspective, the latter are those of the form $\bigl(n, f_2(n)\bigr)$, and the remaining edges correspond to those of form $(x, k)$ to $(x, k - 1)$. We will call the edges of form $\bigl(n, f_2(n)\bigr)$ \defined{jump edges}, and the other edges \defined{predecessor edges}. In our figures, the predecessor edges are \textcolor{cDeemph}{deemphasized}.

The antimonotonicity of $G_{ls2}$ implies that every jump edge of a vertex in the shortest path from some $m$ to $1$ leads into that path again. Assume toward a contradiction there is a jump edge $(z, x)$ such that $z$ is part of the path but $x$ is not. Then the path must contain some other jump edge $(y, w)$ with $w < x < y$, contradicting antimonotonicity. By the isomorphism, it immediately follows that all jump edges from vertices of the certificate pool of $n$ lead into the certificate pool again, and thus do not enlarge its out-neighborhood.

\begin{figure*}
  \centering
  \includegraphics{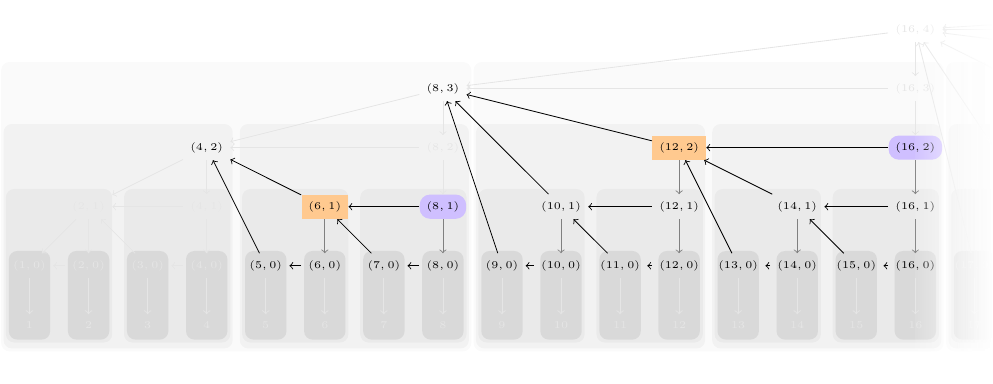}
  \caption{The cores of generations three and four for $SLLS_2$, highlighting their \textcolor{cStart}{starts} and \textcolor{cMiddle}{middles}.}
  \label{fig_slls2_concrete_cores}
\end{figure*}

\begin{figure}
  \centering
  \includegraphics{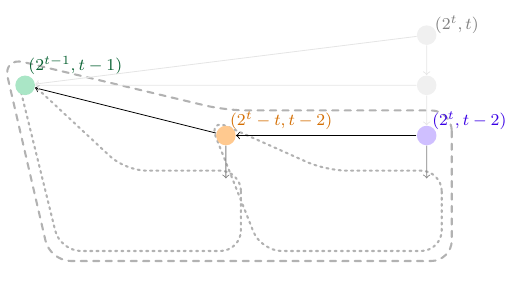}
  \caption{The recursive structure of cores for $SLLS_2$: the core of generation $t$ (dashed) consists of two copies of the core of generation $t - 1$ (dotted).}
  \label{fig_slls2_abstract_cores}
\end{figure}

To analyze the predecessor edges, we define the \defined{core} of generation $t$ as the subgraph of $G_{slls2}$ induced by the longest path from $(2^{t}, t - 2)$ to $(2^{t - 3}, t - 3)$, the \defined{start} of (the core of) $t$ as $(2^t, t - 2)$, and the \defined{middle} of (the core of) $t$ as $(2^t - t, t - 2)$ --- see \cref{fig_slls2_concrete_cores} for some examples. \Cref{fig_slls2_abstract_cores} visualizes the recursive structure of cores: the core of generation $t$ consists of two copies of the core of generation $t - 1$.

This recursive structure allows us to compute the number of predecessor edges on the shortest path from vertebra to vertebra via some $(n, 0)$. We exclude the out-neighbors of the final vertex of the path, as those are already accounted for in our calculation for the size of the path from the vertebra of $t - 1$ to $(1, 0)$.

Let $n \in \N, n > 4$ be of generation $t > 2$, then the vertebra of $t$ is $(2^t, t)$, the \textcolor{cCertPathText}{start of the core is $(2^{t}, t - 2)$}, the \textcolor{cCertText}{middle of the core is $(2^{t} - 2^{t - 1}, t - 2)$}, and the \textcolor{cPrevCoreText}{previous vertebra is $(2^{t - 1}, t - 1)$}. The skip edges of $(2^t, t)$ and $(2^t, t - 1)$ lead out of the current generation, so the path to $(n, 0)$ has to begin with $\bigl((2^t, t), (2^t, t - 1), \textcolor{cCertPathText}{(2^t, t - 2)}\bigr)$, and hence, the predecessor edges of both $(2^t, t)$ and $(2^t, t - 1)$ stay within the path.

\begin{figure*}
  \centering
  \includegraphics{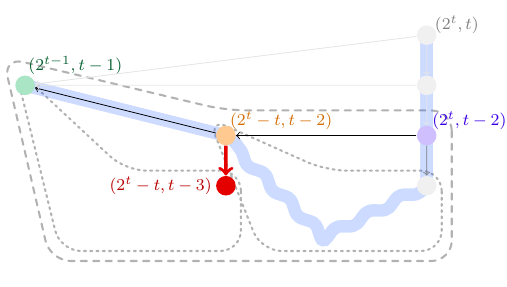}\includegraphics{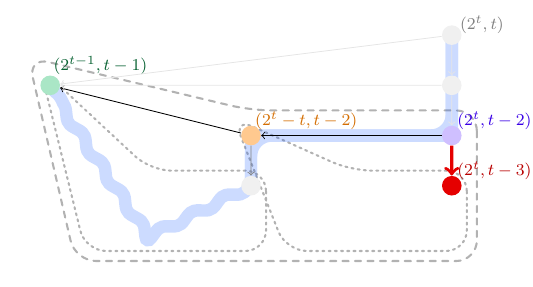}
  \caption{The two possible path shapes in the inductive step. Whether $n$ is in the first or second half of the core, there is exactly \textcolor{myRedText}{one new out-neighbor} of the path compared to the path through the core of the previous generation.}
  \label{fig_slls2_proof}
\end{figure*}

The behavior of the predecessor edge of \textcolor{cCertPathText}{$(2^{t}, t - 2)$} depends on whether $n$ falls into the first or second half of the core (\cref{fig_slls2_proof}). If $n > 2^t - 2^{t - 1}$, the path continues to $(2^t, t - 3)$, so the predecessor edge of \textcolor{cCertPathText}{$(2^{t}, t - 2)$} leads into the path. In this case, however, the path to \textcolor{cPrevCoreText}{$(2^{t - 1}, t - 1)$} ends with $\bigl(\textcolor{cCertText}{(2^{t} - 2^{t - 1}, t - 2)}, \textcolor{cPrevCoreText}{(2^{t - 1}, t - 1)}\bigr)$, so the predecessor edge of \textcolor{cCertText}{$(2^{t} - 2^{t - 1}, t - 2)$} leads outside the path. If $n \leq 2^t - t$, the path continuous from \textcolor{cCertPathText}{$(2^{t}, t - 2)$} to \textcolor{cCertText}{$(2^{t} - 2^{t - 1}, t - 2)$}, so the predecessor edge of \textcolor{cCertPathText}{$(2^{t}, t - 2)$} leads outside the path.

In both cases, the size of the out-neighborhood is one plus the out-neighborhood of the corresponding path in the core of generation $t - 1$. Together with the base case of an out-neighborhood of size two for $t = 2$, induction yields an overall size of $t = \ceil{\log_2(n)}$. Thus, the out-neighborhood of the certificate pool of any $n > 4$ is indeed $2 \cdot \ceil{\log_2(n)}$.

\subsection{Timestamping Rounds}

While not our primary focus, we briefly sketch how the $SLLS_2$ meets the optimal positional certificate size in the setting of timestamping with rounds of bounded length. We let $N$ denote the round length and $T$ denote the generation of $N$; to simplify the presentation we assume that $N$ is a power of two.

\begin{figure*}
  \centering
  \includegraphics{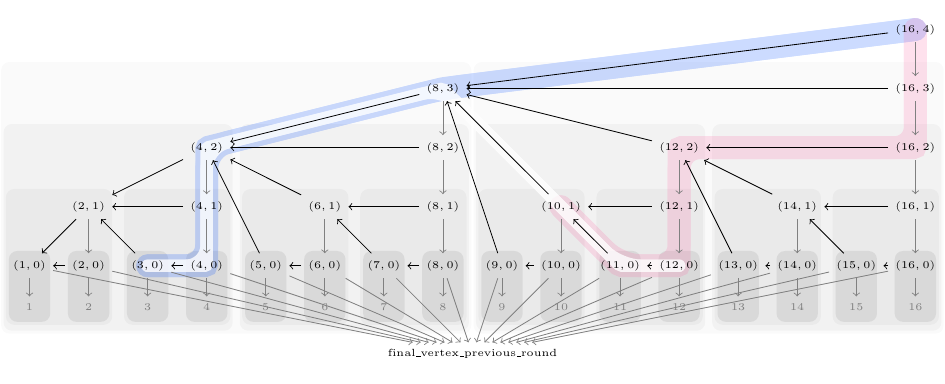}
  \caption{The certificate pools of $3$ and $12$, two numbers from different generations, in a timestamping round of size $N = 16$.}
  \label{fig_pool_bounded_a}
\end{figure*}

\begin{figure*}
  \centering
  \includegraphics{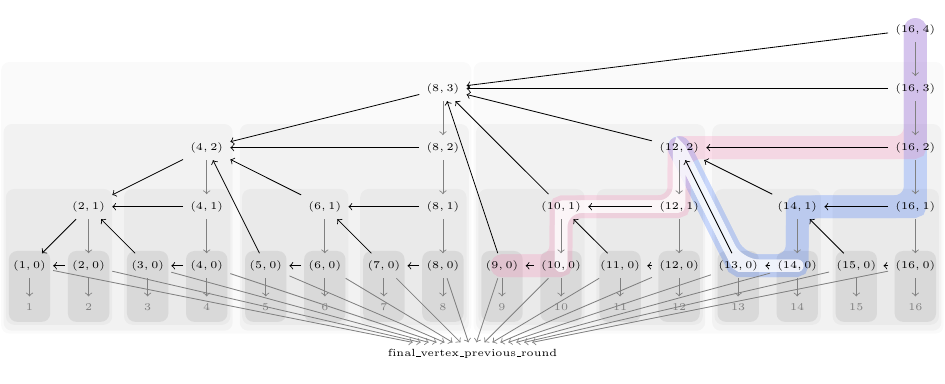}
  \caption{The certificate pools of $10$ and $14$, two numbers from the same generation, in a timestamping round of size $N = 16$.}
  \label{fig_pool_bounded_b}
\end{figure*}

Let $n \leq N$ be of generation $t$. Let $s$ be the vertex just before the vertebra of $t - 1$ on the shortest path from $(n, 0)$ to the vertebra of $t - 1$. Then the certificate pool of $n$ for rounds of length $N$ is the shortest path from $\bigl(N, \log_2(N)\bigr)$ to $s$ via $(n, 0)$. \Cref{fig_pool_bounded_a} and \cref{fig_pool_bounded_b} exemplify how the union of two certificate pools contains the shortest path between their two numbers, whether the numbers are from the same generation or not.

In the previous section, we have shown that the size of the out-neighborhood of the shortest path from the vertebra of $t$ to the vertebra of $t - 1$ via $(n, 0)$ --- excluding the out-neighbors of the vertebra of $t - 1$ --- is $\ceil{\log_2(n)}$. In this new setting, the path stops at $s$ already, so the size of the out-neighborhood increases by one, as the vertebra of $t - 1$ itself becomes part of the out-neighborhood.

If $t = T$, this path is the certificate pool of $n$, and it thus has size $\ceil{t} + 1 = \ceil{\log_2(N)} + 1$. If $t < T$, the certificate pool of $n$ consists of the shortest path from $\bigl(N, \log_2(N)\bigr)$ to the vertebra of $t$, followed by the shortest path from the vertebra of $t$ to the vertebra of $t - 1$ via $(n, 0)$. The latter has size $t + 1$, and the prior is a path of length $T - t$, and each of its vertices increases the size of the out-neighborhood by one, except for the final vertex, whose two out-neighbors are already counted or included in the path from the vertebra of $t$ to the vertebra of $t - 1$ via $(n, 0)$. As such, the total size is $t + 1 + (T - t) = T + 1 = \ceil{\log_2(N)} + 1$ again.
 
When we add an edge from each $(n, 0)$ to the final vertex of the previous timestamping around for inter-round timestamping, we obtan the final positional certificate size: $(\ceil{\log_2(N)} + 2) \cdot k$, where $k$ is the size of an individual hash. This size is provably optimal~\cite{buldas2000optimally}.

\section{Ternary Skip List Scheme}\label{lssl3}

Choosing a base other than $2$ in the definition of a skip-list linking scheme results again in a linking scheme. Of particular interest is base $3$, as this turns out to yield a more efficient construction in the round-less setting, the \defined{ternary skip list linking scheme} (\defined{$SLLS_3$}):

\begin{align*}
  V_{slls3} &\defeq \set{(n, k) \st \text{$n \in \N, k \in \Nz$ and $3^k \divides n$}}\cup \N,\\
  E_{slls3} &\defeq \Bigl\{\bigl((n, k + 1), (n, k)\bigr)\Bigr\}\\
  &\cup \Bigl\{\bigl((n + 3^k, k), (n, \maxpow{3}{n})\bigr)\Bigr\}\\
  &\cup \Bigl\{\bigl((n + 2 \cdot 3^k, k), (n + 3^k, \maxpow{3}{n})\bigr)\Bigr\},\\
  G_{slls3} &\defeq (V_{slls3}, E_{slls3}).
\end{align*}

\begin{figure*}
  \centering
  \includegraphics[width=18cm]{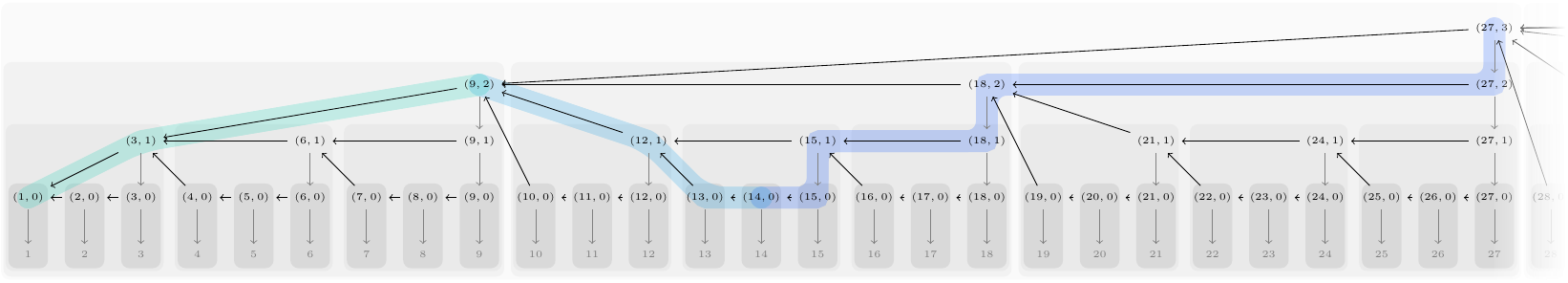}
  \caption{The $SLLS_3$, highlighting the certificate pool of $14$.}
  \label{fig_slls3_basic}
\end{figure*}

\Cref{fig_slls3_basic} depicts the graph.

The remaining discussion is completely analogous to that of the $SLLS_2$.
% Define $\nextv(n)$ as $(n + 1, 0)$ if $\maxpow{3}{n} = 0$ and $(n, 1)$ otherwise. Then define $\gcommit(n) \defeq (n, 0)$, and define $\gcertify(len_s, len_t)$ as the shortest path from $\gcommit(len_t)$ to $\nextv(len_s)$ to obtain a linking scheme.

The $SLLS_3$ (without the sinks) is isomorphic to the \textit{optimal antimonotone linking scheme}~\cite{buldas1998new} (without the sinks), and to the limit of the recursive antimonotone graph product $G_{opt}^{i + 1} \defeq G_{opt}^i \otimes G_{opt}^i \otimes G_{opt}^i \otimes G_1$. 

We define the \defined{generation} of $n$ as $\ceil{\log_3(n)}$. We say $\bigl(t, \log_3(t)\bigr)$ is the \defined{vertebra} of generation $t$, and the set of vertebra of all generations up to and including $t$ is the \defined{spine} of generation $t$. The certificate pool of $n$ is the union of the shortest paths from the vertebra of $t$ to $(n, 0)$, from $(n, 0)$ to the vertebra of $t - 1$, and from that vertebra to $(1, 0)$.

This yields positional certificates of size $3 \cdot \ceil{\log_3(n)}$; the proof is analogous to that for $SLLS_2$. Let $n$ be of generation $t$, then the out-neighborhood of the path from the vertebra of $t - 1$ to $(1, 0)$ has size $t$.

\begin{figure*}
  \centering
  \includegraphics{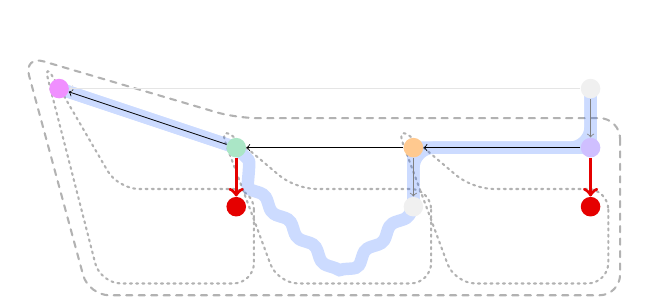}
  \caption{The recursive structure of the cores of $SLLS_3$, each consisting of three copies of the core of the previous generation. No matter in which of the sub-cores a vertex $(n, 0)$ lies, the start vertices of the other two sub-cores increase the size of the out-neighborhood of the certificate pool by two compared to the previous generation.}
  \label{fig_slls3_proof}
\end{figure*}

For the remaining path, we again consider \defined{cores}, this time beginning at $(3^t, t - 2)$ and being split into \textit{three} parts at the vertices $(3^t - 3^{t-1}, t - 2)$ and $(3^t - 2 \cdot 3^{t-1}, t - 2)$. Each of these parts is isomorphic to the core of generation $t - 1$, allowing us to apply induction on $t$ (see \cref{fig_slls3_proof}). The shortest path from the vertebra of $t$ to the vertebra of $t - 1$ via $(n, 0)$ descends into one of the three parts of the core, the predecessor edges of the start vertices of the other two parts lead outside the path. Hence, the size of the out-neighborhood of this path is $2 \cdot t$, for an overall positional certificate size of $3 \cdot t = 3 \cdot \ceil{\log_3(n)}$.

This makes $SLLS_3$ positional certificates about 95 percent the size of $SLLS_2$ positional certificates, as $\frac{3 \cdot \log_3(n)}{2 \cdot \log_2(n)} = \frac{\log_e{8}}{\log_e{9}} \approx 0.9464$. \Cref{fig_sizes} list the positional certificate sizes for all $n$ up to $2^{33}$, to aid with the decision of whether to go with the simpler $SLLS_2$ or the asymptotically more eficient $SLLS_3$. For all $n > 2^{16}$, $3 \cdot \ceil{\log_3(n)}$ is strictly less then $2 \cdot \ceil{\log_2(n)}$.

\begin{table}
  \centering
  \begin{tabular}{| c | c c || c | c c |} 
   \hline
    $\leq$ & $2$ & $3$ & $\leq$ & $2$ & $3$ \\
   \hline\hline
   1 & 1 & 1 & 131072 & 34 & 33 \\
   \cellcolor{cLightGrey} 2 & \cellcolor{cLightGrey} 2 & \cellcolor{cLightGrey} 3 & 177147 & 36 & 33 \\
   3 & 4 & 3 & 262144 & 36 & 36 \\
   \cellcolor{cLightGrey}4 &\cellcolor{cLightGrey} 4 &\cellcolor{cLightGrey} 6 & 524288 & 38 & 36 \\
   8 & 6 & 6 & 531441 & 40 & 36 \\
   9 & 8 & 6 & 1048576 & 40 & 39 \\
   \cellcolor{cLightGrey}16 &\cellcolor{cLightGrey} 8 &\cellcolor{cLightGrey} 9 & 1594323 & 42 & 39 \\
   27 & 10 & 9 & 2097152 & 42 & 42 \\
   \cellcolor{cLightGrey}32 &\cellcolor{cLightGrey} 10 &\cellcolor{cLightGrey} 12 & 4194304 & 44 & 42 \\
   64 & 12 & 12 & 4782969 & 46 & 42 \\
   81 & 14 & 12 & 8388608 & 46 & 45 \\
   \cellcolor{cLightGrey}128 &\cellcolor{cLightGrey} 14 &\cellcolor{cLightGrey} 15 & 14348907 & 48 & 45 \\
   243 & 16 & 15 & 16777216 & 48 & 48 \\
   \cellcolor{cLightGrey}256 &\cellcolor{cLightGrey} 16 &\cellcolor{cLightGrey} 18 & 33554432 & 50 & 48 \\
   512 & 18 & 18 & 43046721 & 52 & 48 \\
   729 & 20 & 18 & 67108864 & 52 & 51 \\
   \cellcolor{cLightGrey}1024 &\cellcolor{cLightGrey} 20 &\cellcolor{cLightGrey} 21 & 129140163 & 54 & 51 \\
   2048 & 22 & 21 & 134217728 & 54 & 54 \\
   2187 & 24 & 21 & 268435456 & 56 & 54 \\
   4096 & 24 & 24 & 387420489 & 58 & 54 \\
   6561 & 26 & 24 & 536870912 & 58 & 57 \\
   \cellcolor{cLightGrey}8192 &\cellcolor{cLightGrey} 26 &\cellcolor{cLightGrey} 27 & 1073741824 & 60 & 57 \\
   16384 & 28 & 27 & 1162261467 & 62 & 57 \\
   19683 & 30 & 27 & 2147483648 & 62 & 60 \\
   32768 & 30 & 30 & 3486784401 & 64 & 60 \\
   59049 & 32 & 30 & 4294967296 & 64 & 63 \\
   \cellcolor{cLightGrey}65536 &\cellcolor{cLightGrey} 32 &\cellcolor{cLightGrey} 33 & 8589934592 & 66 & 63 \\

    \hline
   \end{tabular}
  \caption{The number of vertices in the out-neighborhood of the certificate pools for all $n \leq 2^{33}$ of $SLLS_2$ and $SLLS_3$. For example, for $64 < n \leq 81$, the out-neighborhood has size $14$ for $SLLS_2$ and size $12$ for $SLLS_3$. The highlighted rows are those where $SLLS_2$ is more efficient than $SLLS_3$.}
  \label{fig_sizes}
\end{table}

We can naturally generalize base-3 skip list construction to arbitrary integer bases; the corresponding antimonotone schemes have been studied by Buldas and Laud~\cite{buldas1998new}. Definitions and proofs for positional certificates generalize nicely, they have size $b \cdot \ceil{\log_b(n)}$ for the base-b skiplist scheme. $3 \cdot \ceil{\log_3(n)}$ is the slowest-growing function of this family, making the most efficient scheme $SLLS_3$ and the simplest scheme $SLLS_2$ the only two members of this family of practical interest.

\section{Discussion}\label{discussion}

The $SLLS_2$ achieves the same positional certificate size as threaded authentication trees~\cite{buldas2000optimally}, hypercore~\cite{ogden2017dat}, and certificate transparency logs~\cite{laurie2014certificate}, the best previously known schemes in that respect. Unlike these schemes, it does so with a underlying graph of linear size: every vertex has out-degree at most $2$, and the number of vertices of the subgraph for a sequence of length $n$ is at most $3n$. Furthermore, the digest vertex of every sequence item is the direct parent of the sequence item, whereas the distance between them can be logarithmic in hypercore and certificate transparency logs.

\begin{figure*}
  \centering
  \includegraphics[width=18cm]{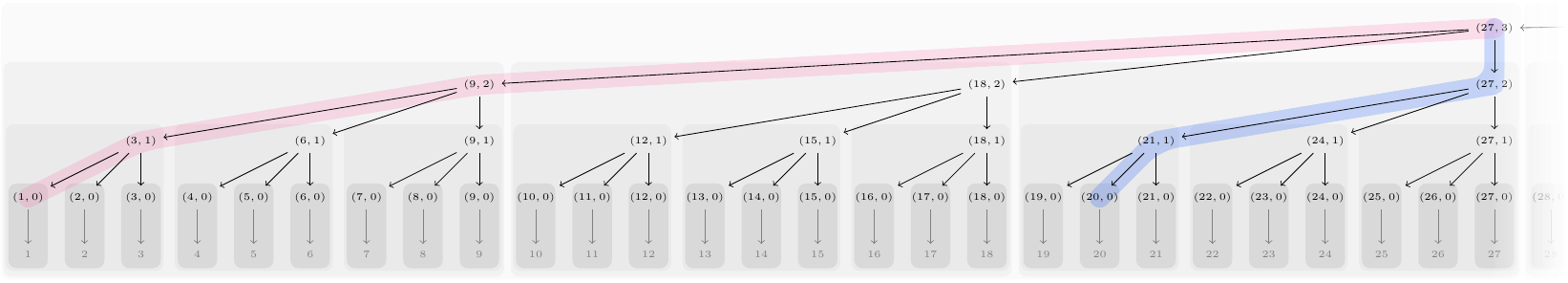}
  \caption{Basing hypercore-like or transparency-log-like schemes off ternary trees is inefficient: the paths from a root to the two leaves each have 2 out-neighbors per vertex, for an overall out-neighborhood size of $4 \cdot \ceil{\log_3(n)} - 1$ rather than the desired $3 \cdot \ceil{\log_3(n)}$.}
  \label{fig_hypercore3nope}
\end{figure*}

\begin{figure*}
  \centering
  \includegraphics[width=18cm]{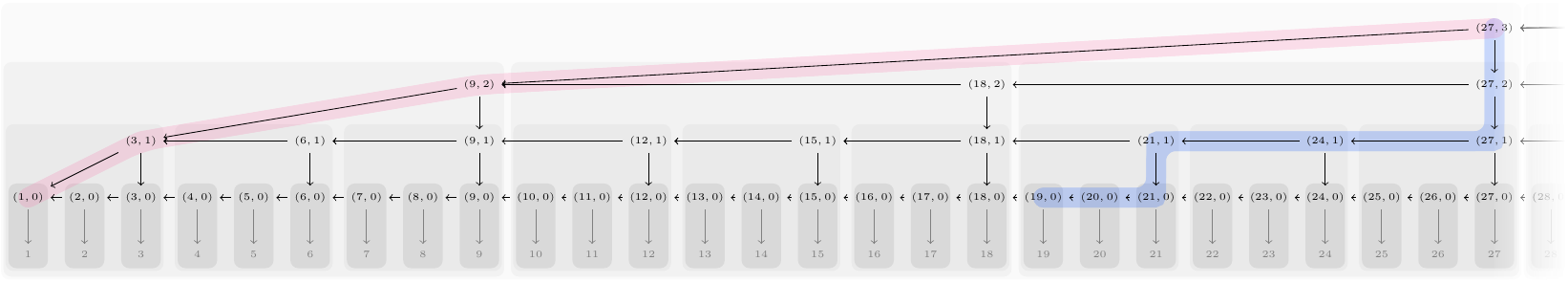}
  \caption{A conventional ternary skip list is inefficient as well: while the path to $(1, 0)$ only contributes one out-neighbor per vertex, the other path contributes three out-neighbors per level of the skip list.}
  \label{fig_hypercore3nope2}
\end{figure*}

As such, the $SLLS_2$ strictly outperforms all existing schemes that achieve positional certificates of size $2 \cdot \ceil{\log_2(n)}$. The $SLLS_3$ then further improves upon the positional certificate size, it is the first scheme to beat the positional certificate size of $2 \cdot \ceil{\log_2(n)}$. It is worth noting that simply adapting hypercore or transparency logs to a ternary rather than binary tree structure does \textit{not} result in positional certificates of size $3 \cdot \ceil{\log_3(n)}$ (see \cref{fig_hypercore3nope} and \cref{fig_hypercore3nope2}). The skip-list-with-a-twist is qualitatively different from tree structures in that regard.

For timestamping with rounds of bounded length, $SLLS_2$ is the first scheme to achieve minimal structural certificates with a graph of linear size.

\Cref{fig_summary} takes the (competitive) schemes surveyed in \cite{meyer2023prefix} and compares them to our new schemes. The picture is quite clear: if any new sequence item must add no more than $\complexity{1}$ of metadata, use the optimal antimonotone linking scheme. In every other case, use the $SLLS_2$ (if using rounds of bounded length) or the $SLLS_3$.

\begin{table*}
  \centering
  \begin{tabular}{| c || c c c c |} 
   \hline
    & Linear & Simple Antimonotone & Optimal Antimonotone & Threaded Authentication \\
   \hline\hline
   Positional Certificate & $n \cdot k$ & $(5 \cdot \floor{\log_2(n)} - 3) \cdot k$ & $(7 \cdot \floor{\log_3(2n)} - 4) \cdot k$ & $2 \cdot \ceil{\log_2(n)} \cdot k$ \\ 
   Certificate Validation & $\complexity{certsize}$ & $\complexity{certsize}$ & $\complexity{certsize}$ & $\complexity{certsize \cdot \log(certsize)}$ \\
   Edges Amortized & $\complexity{n}$ & $\complexity{n}$ & $\complexity{n}$ & $\complexity{n \cdot \log(n)}$ \\
   Edges Worst Case & $\complexity{1}$ & $\complexity{1}$ & $\complexity{1}$ & $\complexity{\log(n)}$  \\
   Vertices Amortized & $\complexity{n}$ & $\complexity{n}$ & $\complexity{n}$ & $\complexity{n}$ \\
   Vertices Worst Case & $\complexity{1}$ & $\complexity{1}$ & $\complexity{1}$ & $\complexity{\log(n)}$ \\
   Identifier Amortized & $\complexity{1}$ & $\complexity{1}$ & $\complexity{1}$ & $\complexity{1}$ \\
   Identifier Worst Case & $\complexity{1}$& $\complexity{1}$ & $\complexity{1}$ & $\complexity{1}$ \\
   Digest Pool & $1$ & $\floor{\log_2(n)}$ & $\floor{\log_3(2n)}$ & $\floor{\log_2(n)}$ \\ 
   \hline
    \hline
     & Hypercore & Transparency Log & $SLLS_2$ & $SLLS_3$ \\
    \hline\hline
    Positional Certificate & $2 \cdot \ceil{\log_2(n)} \cdot k$ & $2 \cdot \ceil{\log_2(n)} \cdot k$ & $2 \cdot \ceil{\log_2(n)} \cdot k$ & $3 \cdot \ceil{\log_3(n)} \cdot k$ \\ 
    Certificate Validation & $\complexity{certsize}$ & $\complexity{certsize}$ & $\complexity{certsize}$ & $\complexity{certsize}$ \\
    Edges Amortized & $\complexity{n \cdot \log(n)}$ & $\complexity{n \cdot \log(n)}$ & $\complexity{n}$ & $\complexity{n}$ \\
    Edges Worst Case & $\complexity{\log(n)}$ & $\complexity{\log(n)}$ & $\complexity{1}$ & $\complexity{1}$ \\
    Vertices Amortized & $\complexity{n}$ & $\complexity{n \cdot \log(n)}$ & $\complexity{n}$ & $\complexity{n}$ \\
    Vertices Worst Case & $\complexity{\log(n)}$ & $\complexity{\log(n)}$ & $\complexity{\log(n)}$ & $\complexity{\log(n)}$ \\
    Identifier Amortized & $\complexity{1}$ & $\complexity{1}$ & $\complexity{1}$ & $\complexity{1}$ \\
    Identifier Worst Case & $\complexity{\log(n)}$ & $\complexity{\log(n)}$ & $\complexity{1}$ & $\complexity{1}$ \\
    Digest Pool & $\floor{\log_2(n)}$ & $\floor{\log_2(n)}$ & $\floor{\log_3(2n)}$ & $\floor{\log_2(n)}$ \\ 
    \hline
   \end{tabular}
  \caption{Summary of prior competitive schemes and our new schemes, according to the complexity criteria layed out in~\cite{meyer2023prefix}.}
  \label{fig_summary}
\end{table*}

\section{Conclusion}\label{conclusion}

We have provided prefix authentication schemes that outperform the previous state of the art. We hope that future endeavors around certificate transparency in particular will adopt our schemes.

We did not consider questions of optimality, leaving open whether schemes with smaller positional certificates exist.

%-------------------------------------------------------------------------------
\bibliographystyle{plain}
\bibliography{\jobname}

%%%%%%%%%%%%%%%%%%%%%%%%%%%%%%%%%%%%%%%%%%%%%%%%%%%%%%%%%%%%%%%%%%%%%%%%%%%%%%%%
\end{document}